%% file: main.tex
\documentclass[a4paper, 10pt, conference]{ieeeconf} 
\IEEEoverridecommandlockouts                       
\usepackage{graphicx}
\usepackage{amsmath,mathtools}
\usepackage{xcolor}
\usepackage{fancyvrb}
\usepackage{tikz}
\usetikzlibrary{shapes.geometric, arrows}
\usetikzlibrary{shapes, arrows}
\usepackage{amssymb}
\usepackage{empheq}

\definecolor{dark}{rgb}{0.2,0.2,0}
\definecolor{lightred}{rgb}{1,0.3,0.3}
\definecolor{darkyellow}{rgb}{0.3,0.3,0}
\definecolor{mediumgreen}{rgb}{0,0.6,0}
\definecolor{darkgreen}{rgb}{0,0.3,0}
\definecolor{lightblue}{rgb}{0.6,0.7,1}
\definecolor{colgate}{rgb}{0.1,0.1,0}

\newcommand{\fig}[1]{Fig.~\ref{#1}}

\newcommand{\fm}[1]{\mbox{$#1$}}

\newcommand{\gand}{{\color{colgate}\tt{}AND}}

\newcommand{\gor}{{\color{colgate}\tt{}OR}}

\newcommand{\gxor}{{\color{colgate}\tt{}XOR}}
\newcommand{\gnot}{{\color{colgate}\tt{}NOT}}

\newcommand\Smallmath[2]{#1{\raisebox{\dimexpr \fontdimen 22 \textfont 2
      - \fontdimen 22 \scriptscriptfont 2 \relax}{$\scriptstyle{}#2$}}}
\newcommand\smallmath[2]{#1{\raisebox{\dimexpr \fontdimen 22 \textfont 2
      - \fontdimen 22 \scriptfont 2 \relax}{$\scriptstyle{}#2$}}}
\newcommand\soplus{\;\!{\smallmath\mathbin\oplus}\;\!}
\newcommand\Soplus{\;\!{\Smallmath\mathbin\oplus}\;\!}
\newcommand\samp{\texttt{\&}}
\newcommand\spar{\,{\Smallmath\mathbin\parallel}\,}

\input{tikzini}

\title{\LARGE \bf
Modular Multiplication without Carry Propagation \\ 
{\large Algorithm Description}
}

\author{Oleg Mazonka\\
{\it New York University Abu Dhabi, UAE}, 2021}

\begin{document}

\maketitle
\thispagestyle{empty}
\pagestyle{empty}

\begin{abstract}

This paper describes a sufficiently simple modular multiplication algorithm, which uses only carry-save addition with bit inspection Boolean logic and without number comparison or carry propagation.

\end{abstract}


\input{00-body.tex}

\end{document}

%% file: tikzini.tex
\tikzstyle{io} = [rectangle, rounded corners, minimum width=3cm, minimum height=0.5cm,text centered, draw=black, fill=red!10]

\tikzstyle{shd} = [trapezium, trapezium left angle=70, trapezium right angle=70, text width=2.1cm, minimum height=0.5cm, text centered, draw=black, fill=green!20]
\tikzstyle{shu} = [trapezium, trapezium left angle=110, trapezium right angle=110, text width=2.1cm, minimum height=0.5cm, text centered, draw=black, fill=green!20]

\tikzstyle{process} = [rectangle, minimum width=1cm, minimum height=1.75em, text centered, draw=black,
text width=2cm,
fill=orange!20]

\tikzstyle{decision} = [diamond, minimum width=1cm, minimum height=0.5cm, text centered, draw=black, fill=green!20]
\tikzstyle{arrow} = [thick,->,>=stealth]
\tikzstyle{darr} = [double,->,>=stealth]

\tikzstyle{block} = [draw, fill=blue!20, rectangle, minimum height=2em, minimum width=2em]

\tikzstyle{rule} = [draw, fill=cyan!20, rectangle, minimum height=1.5em, text width=0.9cm, node distance=0.75cm, align=center]

\tikzstyle{ruleDots} = [draw, fill=cyan!10, rectangle, minimum height=1.5em, text width=0.9cm, node distance=0.75cm, align=center, draw=white, rounded corners=0.3cm]

\tikzstyle{topup} = [draw, fill=magenta!20, rectangle, minimum height=1.5em, minimum width=2em, node distance=0.75cm]

\tikzstyle{dot} = [draw,circle,node distance=1cm,inner sep=0pt,minimum size=5pt]
\tikzstyle{sum} = [draw, fill=orange!20, circle, node distance=1cm]

\tikzstyle{point} = [coordinate, node distance=1cm]
\tikzstyle{pinstyle} = [pin edge={to-,thin,black}]
\tikzstyle{none} = [draw=none, circle, node distance=1cm,,inner sep=0pt,minimum size=0pt]
\tikzstyle{mux} = [trapezium, trapezium left angle=70, trapezium right angle=70, minimum width=1cm, minimum height=0.5cm, text centered, draw=black, fill=green!30]


%% file: 00-body.tex
\input{10-intro}

\input{20-overview}
\input{30-loop}

\input{40-shrink}
\input{45-cyc}

\input{50-squee}

\section{Acknowledgements}
I would like to thank my colleagues Michail Maniatakos and Eduardo Chielle for valuable recommendations and improvements to the paper.

\section{Final remarks}

The purpose of this paper is to present the mathematical solution to the problem. This work {\it does not} compare the performance to other algorithms (such as \cite{brick}, \cite{koch}); and does not discuss applications or hardware implementation.
Due to no carry propagation, the presented algorithm can be extended to process either long number operations or several shorter in parallel on the same hardware, i.e. reusing operational elements for different length of the operands. Also in chained computation, it can be extended to allow input in the form of \fm{(P,Q)}. The idea can be extended to use 7-to-3 carry-save adders with higher Radix processing.
The presented algorithm has been developed without awareness of the plethora of existing carry-save based algorithms. 
The next step will be to analyse and compare to the other modular multiplication algorithms and possibly extend and optimize it for specific applications.

\bibliographystyle{plain}
\bibliography{ref}

%% file: 10-intro.tex
\section*{Notation and operations}
In this paper the following notation is used.
Numbers represented as big letters, e.g. $A,B$. Their corresponding bits of binary expansion represented by small indexed letters, e.g. $a_0, b_i$. Index $0$ specifies the least significant bit.
Letters $n$ and $k$ are sizes of the numbers in bits. Bars $|\cdot|$ denote {\it bit-length}: so $|A|=n$ means that $A$ has exactly $n$ bits. Operation $[\cdot]_R$ is modular reduction by $R$ such that  \fm{A\!\equiv\![A]_R\bmod R} and \fm{[A]_R\!<\!R}. Operation
\fm{\lfloor\cdot\rfloor} is floor function.
Operations $(+,-)$ are arithmetic addition and subtraction. Arithmetic multiplication is {\it implied} between numbers, e.g. $AB$. 
Boolean multiplication, \fm{\gand} gate, is implied between bits, e.g. \fm{ab=\gand(a,b)}. Other Boolean operations are explicit: $a \soplus b\!\equiv\!\gxor(a,b)$ and $a\spar b\!\equiv\!\gor(a,b)$.
Boolean negation is denoted with a bar: \fm{\overline{a}\!\equiv\!\gnot{}(a)}.
Boolean operations can be used between numbers in parallel bit-to-bit manner.
Explicit bit-by-bit Boolean \fm{\gand{}} as $A\samp B$ can be used between numbers. Symmetric 2-out-of-3 function is denoted as:
\begin{equation*}
    (a,b,c)_2=ab\spar bc\spar ac=ab\spar c(a\spar b)
\end{equation*}
Further in the text, terms {\it top bit} and {\it most significant bit} mean the same. The term {\it top bits} means several high order bits including the top bit.

\section{Introduction}
\subsection{Background}
Can we do modular multiplication of two numbers without number comparison, conventional arithmetic subtraction or addition; or any other elementary operations that depend on the size of the operands of multiplication?
This paper describes one possible and sufficiently simple algorithm.

Carry propagation is necessary in normal arithmetic processing numbers. For example, in 
\begin{equation*}
15+1=1111_b+1_b=10000_b=16    
\end{equation*}
in binary representation the most significant bit of the result depends on the least significant bits of the arguments. In order to obtain the result, the addition algorithm must propagate the bit value through the whole length of the number; through 4 bit positions in the above example. Therefore, number bit-length affects the 
length of the computation sequence -- it is impossible to get the result of the top bit without computing first all lower bit positions.

There are a few ways to rewrite a sum of two or three numbers:
\begin{equation*}
\begin{split}
    A+B&=(A\Soplus B)+2(A\samp B) \\
    A+B&=(A{\spar} B)+(A\samp B) \\
    A+B+C&=(A\spar B\spar C)+(A,B,C)_2+(A\samp B\samp C) \\
    A+B+C&=(A\Soplus{}B\Soplus{}C)+2(A,B,C)_2
\end{split}
\end{equation*}
The first equation corresponds to carry propagation methods -- the equation is applied in iterations until the last term \fm{(A\samp B)} is zero. The second and the third equations are {\it top-up}, can be used to reorganize the sum so that the operands are sorted by value. This top-up operation is used further in the presented algorithm. The last equation is a carry-save addition with two useful properties: 1) bit operations can be done independently and in parallel for all bit positions; and 2) the sum of 3 numbers is reduced to a sum of 2 numbers. Hence, a long sequence of additions can be done more efficiently using this trick, since carry-save operation does not depend on the length of the operand numbers.

Normal (non modular) multiplication requires many additions. Carry-save addition can naturally be used, so carry propagated addition is done only once at the very end to obtain the result as one number. In modular multiplication it is not obvious how to use carry-save because it requires reduction operation, that requires comparison, and that, in turn, requires subtraction with carry propagation. 
The aim of this work is to perform modular multiplication while avoiding carry propagating addition, subtraction and number comparison.

\subsection{Contribution}
The algorithm\footnote{Internally called IM1C - Interleaved Modular 1-bit-Radix Carry-save.} for modular multiplication \fm{[AB]_R}, where $R$ is modulus, outputs a pair of numbers \fm{(P,Q)} such that \fm{P,Q\!<\!R} and \fm{P\!+\!Q\!\equiv{}\!AB\bmod{R}}. The algorithm uses only carry-save adders and fixed Boolean logic operations of $O(1)$ complexity. 

\subsection{Limitations}

Along with $A,B,R$ input the algorithm requires five precomputed values of $R$, e.g. $[2^{n+1}]_R$. Some of them are hard to compute in carry-save only mode in $O(1)$ number of steps. This makes the algorithm less efficient when $R$ is changing from one multiplication to another.

The algorithm outputs two numbers instead of one. At the time of writing, it not known if producing the result as one final number is possible, given the constrains of using only carry-save addition, fixed Boolean logic of $O(1)$ complexity, and final reduction in $O(1)$ steps. If one number is necessary as the result, an extra addition with carry propagation is needed.

%% file: 20-overview.tex
\section{Algorithm overview}


\subsection{Interleaved multiplication}

The algorithm presented in this paper follows the idea of the classical interleaved modular multiplication \cite{blakley}, which is briefly described below.
The expression, seen as {\it Horner's scheme} in powers of 2 expansion:
\begin{equation*}
    AB=2(2(...2(2a_{k-1}B+a_{k-2}B)+...+a_2B)+a_1B)+a_0B
\end{equation*}
gives a direct way to multiply two numbers $A$ and $B$ modulo $R$, where $k$ is number of bits (binary digits) of each these three numbers. In each iteration the {\it accumulator} (holding the result value) is multiplied by 2; added $a_iB$; and reduced by $R$. In carry-save mode the reduction step is problematic because comparison between the accumulator and the modulus is impossible. The solution presented here is to drop high bits of the accumulator and compensate the accumulator by adding a specific value in such way that the result remains valid.

\subsection{Conditions and notations}
Let's define the working size of the algorithm as $n$, that is the algorithm can do multiplications up to $n$-bit numbers. Let us call two values of the {\it accumulator as $P$ and $Q$}. The current version of the algorithm requires $P,Q$ to be 1 bit larger than the working size: \mbox{$|P|\!=\!|Q|\!=\!n\!+\!1$}. Assume also \mbox{$A,B\!<\!R$}, \mbox{$n\!>\!2$} and \mbox{$n\!\ge\!k$}, where $k$ is determined by \mbox{$2^{k-1}\!\le \!R\!<\!2^k$}. Therefore, \mbox{$|A|\!=\!|B|\!=\!|R|\!=\!k$}.
To simplify expressions in the future, denote $2^k$ as $\beta$: \mbox{$\beta=2^k$}.

\noindent{\bf Input:} $n,k,A,B,\{\!R\}$.

\noindent $\{\!R\}$ is a set of 5 precomputed values of $R$ and the bit next to the most significant bit of $R$:
\begin{align*}
&\{\!R\}\equiv\{R_n, R_m, R_x(i),r_{k-2}\} && \makebox[0.5cm]{} \\
& R_x(i)=\{0,R_1,R_2,R_3\} &&
\end{align*}
\vspace*{-0.75cm}
\begin{align*}
& R_n=[\beta]_R &
& R_m=[{3}\beta/4]_R\\
& R_1=[2\beta]_R &
& R_2=[4\beta]_R \\
& R_3=[6\beta]_R &
& r_{k-2}\!=\!\lfloor{R/2^{k-2}}\rfloor-2
\end{align*}

\noindent{\bf Output:} $(P,Q)$ such that \fm{P\!<\!R} and \fm{Q\!<\!R}, and
\mbox{$P\!+\!Q\!\equiv\!AB \bmod{R}$}.

\subsection{Overall picture}
\fig{f:diag_ov} shows the algorithm overall diagram. 
{\it Main Loop}, {\it Shrink}, and {\it Squeeze} modules are three sequential steps performing computation. They work on the assumption that the most significant bit of $R$ is 1, i.e. \fm{2^{n-1}\!\le \!R\!<\!2^n} (note $n$ instead of $k$).
To accommodate this condition we shift left (normally assuming the top bit to be in the leftmost position) by \fm{n\!-\!k} bit positions (same as multiply by $2^{n-k}$) values $B,R,\{\!R\}$ in {\it Shift-left} step. Accordingly before the output we shift right ({\it Shift-right}) by the same number of bits (division by $2^{n-k}$) output values $P$ and $Q$.

Main loop module executes interleaved iterations over all bits of $A$ in $k$ cycles. Its output is a pair $(P,Q)$ of size \fm{n\!+\!1}, hence, in ranges \mbox{$0\!\le\!P\!<\!2\beta$} and \mbox{$0\!\le\!Q\!<\!2\beta$}. Shrink module reduces values $(P,Q)$ so that their sizes are $n$ and the ranges are \mbox{$0\!\le\!P\!<\!\beta$} and \mbox{$0\!\le\!Q\!<\!\beta$}, and \fm{P\samp Q\!<\!\beta/2}. Finally, Squeeze module does further reduction so the both values $\mbox{$P,Q\!<\!R$}$.

Loop module uses $A$, $B$, and $R_x$ values. Also it uses value $k$ as the number of iterations. Shrink module uses $R_1$ and $R_n$. Squeeze module uses $R_n$, $R_m$, and $r_{k-2}$.

\input{diag2_ov}

%% file: diag2_ov.tex
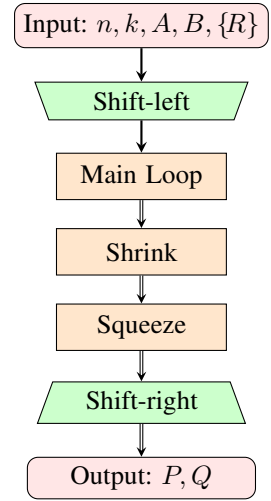
\begin{figure}[t]
\begin{center}
\begin{tikzpicture}[node distance=1cm]
\node (in1) [io] {Input: $n,k,A,B,\{\!R\}$};
\node (shu) [shu, below of=in1] {Shift-left};
\node (pro1) [process, below of=shu] {Main Loop};
\node (pro2) [process, below of=pro1] {Shrink};
\node (pro3) [process, below of=pro2] {Squeeze};
\node (shd) [shd, below of=pro3] {Shift-right};
\node (out1) [io, below of=shd] {Output: $P,Q$};

\draw [arrow] (in1) -- (shu);
\draw [arrow] (shu) -- (pro1);
\draw [darr] (pro1) -- (pro2);
\draw [darr] (pro2) -- (pro3);
\draw [darr] (pro3) -- (shd);
\draw [darr] (shd) -- (out1);
\end{tikzpicture}
\end{center}

\caption{Overall diagram of the algorithm.}
\label{f:diag_ov}
\end{figure}

%% file: 30-loop.tex
\section{Main loop Module}

\input{diag4_lo}

\subsection{Computation flow}

Loop module is the main and most critical part of the  algorithm. It executes the iterations over all bits of $A$ starting from bit \fm{k\!-\!1} and going down to $0$. This is the only place where $k$ is used. From now on it is assumed that \fm{k\!=\!n} because Shift-left has been applied. Correspondingly \fm{\beta\!=\!2^n} in the text below.

As shown in \fig{f:diag_lo} there are two additions $({\Sigma})$, {\it Loop Control Unit} ({LCU}), and the multiplexer producing $R_y$ reduction value. First addition adds to the  accumulator, shifted left by 1 bit position, the next value of $B$ and outputs a pair $(S,C)$:
\begin{equation*}
    (S,C) = (2P,2Q)+a_iB
\end{equation*}
The second addition adds the reduction value $R_y$
\begin{equation*}
    (P,Q) \gets (P',Q') = (S,C)+R_y
\end{equation*}
Note, that both additions are done in carry-save mode in \mbox{$n\!+\!1$} sized registers discarding all overflowing bits.
Discarded bits can be seen as subtraction of some number $F$.
Value $R_y$ must match $F$ to make it valid reduction by $R$:
\begin{equation*}
    R_y - 2\beta F = 0 \bmod R
\end{equation*}
In the above equation $R_y$ is a value we arithmetically add. The second term \mbox{$2\beta F\!\ge\! R_y$} is the value we arithmetically subtract from the accumulator, effectively making reduction.

\subsection{Bit analysis in LCU}
\label{ss:lcu}

The value $F$ has two important properties: 1) it does not depend on $R_y$, i.e. $R_y$ bits do not propagate to the overflow bits; and 2) \mbox{$F\!<\!4$}, i.e. has only two bits $f_1$ and $f_0$: \mbox{$F\!=\!2f_1+f_0$}.
The first property breaks the circular dependency; and the second makes the algorithm requirements and computation simple.

To understand why $F$ has these properties consider the following example. Let $n=4$. The additions can schematically be represented by the worksheet:
\input{t_bits}
The first line is $P$ and the second is $Q$, both are left shifted by 1 bit. The third line represents $a_iB$. The full result of the addition is all the bits $s$ and $c$ on the fourth and fifth lines. Their values are defined by expressions shown below in subsection \ref{ss:cs}.
Note, that the bits on the left side of the vertical line ($p_4$,~$q_4$,~$s_5$,~etc) do not participate in computation because we use only 5-bit carry-save adder.
After the lines with $s$ and $c$ bits, we add $R_y$ value shown as four $r$ bits resulting in new values of $P$ and $Q$, with bits $p_5$, $p_6$, $q_5$, and $q_6$ being overflown. Finally, these overflown bits form the value $F$. 

First observation is that neither $p_5$ nor $q_5$ depend on $r_3$. The highest bit that depends on $r_3$ is $q_4$ and $q_4$ position remains within the \mbox{$n\!+\!1$} bounds. Therefore $F$ does not depend on value $R_y$.
Second observation is that \mbox{$s_5c_5\!=\!0$} and \mbox{$s_4c_4\!=\!0$}, hence the resulting bits satisfy the following both conditions:
\begin{equation*}
\begin{split}
& p_6q_6=c_5s_5c_4=0 \\
& p_5q_5(p_6\spar{}q_6)=(s_5\Soplus{}c_4)s_4c_3(c_5\spar{}s_5c_4)=0
\end{split}
\end{equation*}
which imply no carry to the position above $f_2$ and therefore the sum of two numbers
\fm{(2p_6\!+\!p_5)} and \fm{(2q_6\!+\!q_5)}, which is equal to $F$, has only 2 bits; hence, \fm{F\!<\!4}.

Direct derivation of bit values $f_1$ and $f_0$ gives the control logic for selecting $R_y$:
\begin{align*}
& s_4 = p_3 \soplus q_3 &
& s_5 = p_4 \soplus q_4\\
& c_3 = (p_2, q_2, b_3)_2 &
& c_4 = p_3 q_3\\
& c_5 = p_4 q_4 &
& q_5 = s_4 c_3\\
& f_0 = q_5 \soplus s_5 \soplus c_4 &
& f_1 = c_5 \soplus (s_5, c_4, q_5)_2
\end{align*}
This logic is implemented in LCU.
Finally, the multiplexer selects one correct value from the array of four: zero and three precomputed values (\mbox{$F\!=\!1,2,3$}): 
\begin{equation*}
    R_y = R_x(F) = [2\beta F]_R
\end{equation*}
In this way, when computing $F$, LCU inspects 7 bits: the three highest bits of $P$ and $Q$ and the most significant bit of $a_iB$. Then the multiplexer selects the appropriate reduction value $R_y$.

\subsection{Carry-save adder}
\label{ss:cs}
Carry-save adder, used in this algorithm, 
of size $m$ with three inputs \fm{(X,Y,Z)} and two outputs \fm{(S,C)}
is a standard carry-save operation defined by the functions:
\begin{equation*}
\begin{split}
&S = X\Soplus{}Y\Soplus{}Z \\
&C = 2(X,Y,Z)_2 \bmod {2^m}
\end{split}
\end{equation*}
The reduction on the carry $C$, erasing the top bit after the shift, is necessary for the correct behaviour of the algorithm.

%% file: diag4_lo.tex
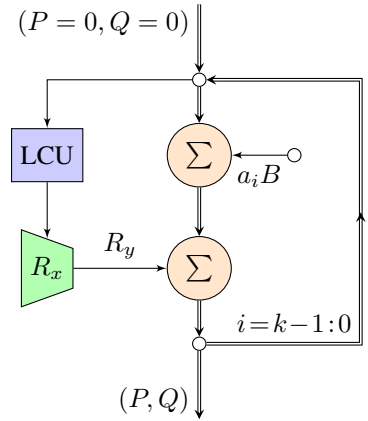
\begin{figure}[t]

\begin{center}

\begin{tikzpicture}[auto, node distance=2cm,>=latex']
\node [point] (input){};
\node [dot, below of=input] (start) {};
\node [sum, below of=start] (feed){$\sum$};
\node [dot, right of=feed, node distance=1.25cm] (aib) {};
\node [block, left of=feed] (lcu){LCU};
\node [sum, below of=feed, node distance=1.5cm] (red){$\sum$};
\draw [darr] (feed) -- node[name=mid]{}(red);

\node [mux, left of=red, rotate=-90](mux){\rotatebox{90} {$R_x$}};
\node [dot, below of=red] (end) {};
\node [point, below of=end, name=out] {};
\node [none, right of=mid, node distance=2cm](arrup){};
\node [none, below of=arrup, node distance=0.06cm](arrud){};

\draw [darr] (input) -- node[anchor=east,yshift=0.2cm] {${(P=0,Q=0)}$} (start);
\draw [darr] (start) -- (feed);
\draw [->] (start) -| (lcu);
\draw [->] (lcu) -- (mux);
\draw [->] (mux) -- node {$R_y$}(red);
\draw [darr] (red) -- (end);
\draw [darr] (end) -- node [anchor=east,yshift=-0.2cm] {$(P,Q)$}(out);

\draw [->] (aib) -- node{$a_iB$}(feed);
\draw [darr] (arrud) |- (start);
\draw [darr] (end) -| node[anchor=south east]{$i\!=\!k\!-\!1\!:\!0$}(arrup);

\end{tikzpicture}

\end{center}

\caption{Loop module consists of two  carry-save adders of size \mbox{$n\!+\!1$}, LCU (Loop Control Unit), and the Multiplexer. LCU takes 7 bits as input and outputs 2 bits. The Multiplexer selects one out four values: 0, $R_1$, $R_2$, $R_3$.}
\label{f:diag_lo}
\end{figure}

%% file: t_bits.tex
\setlength{\tabcolsep}{3pt} 
\renewcommand{\arraystretch}{1} 

\begin{center}
\begin{tabular}{llllllll}
 &  & \multicolumn{1}{l|}{$p_4$} & $p_3$ & $p_2$ & $p_1$ & $p_0$ & 0 \\
 &  & \multicolumn{1}{l|}{$q_4$} & $q_3$ & $q_2$ & $q_1$ & $q_0$ & 0 \\
 &  & \multicolumn{1}{l|}{} &  & $b_3$ & $b_2$ & $b_1$ & $b_0$ \\ \cline{3-8} 
 &  & \multicolumn{1}{l|}{$s_5$} & $s_4$ & \multicolumn{1}{l|}{$s_3$} & $s_2$ & $s_1$ & $s_0$ \\ \cline{5-5}
 & $c_5$ & \multicolumn{1}{l|}{$c_4$} & \multicolumn{1}{l|}{$c_3$} & $c_2$ & $c_1$ & $c_0$ & 0 \\
 &  & \multicolumn{1}{l|}{} &  & $r_3$ & $r_2$ & $r_1$ & $r_0$ \\ \cline{2-8} 
 & $p_6$ & \multicolumn{1}{l|}{$p_5$} & $p_4$ & $p_3$ & $p_2$ & $p_1$ & $p_0$ \\ \cline{4-4}
 & $q_6$ & \multicolumn{1}{l|}{$q_5$} & $q_4$ & $q_3$ & $q_2$ & $q_1$ & $q_0$ \\ \cline{1-3}
0 & $f_1$ & $f_0$ &  &  &  &  & 
\end{tabular}
\end{center}

%% file: 40-shrink.tex
\section{Shrink module}

\subsection{Components}

Shrink module, schematically depicted in \fig{f:diag_sh}, performs the reduction of the accumulator by erasing the top bits of $P$ and $Q$ effectively shrinking their size by one bit from \fm{n\!+\!1} to $n$; as well as ensuring that \fm{P\samp{}Q\!<\!\beta/2}.
It runs in cycles up to three times following the logic:
\begin{enumerate}
    \item {\it Top-up} moves top bits between $P$ and $Q$;
    \item {\it Shrink Control Unit} (SCU) analyses a few top bits of $P$ and $Q$ and selects one of four rules to apply;
    \item Each rule executes one carry-save summation, and clears some bits if necessary.
\end{enumerate}
For the sake of simplicity and without losing generality let us use 4-based indices as shown in the example in Section~\ref{ss:lcu} instead of $n$-based. So $p_4$ is $p_n$, $q_3$ is $q_{n-1}$ and so on.

{Top-up} operation consists of changing two top bits in $Q$ to $P$ as:
\begin{align*}
    &p_4' = p_4\spar{}q_4 & &q_4' = p_4q_4\\
    &p_3' = p_3\spar{}q_3 & &q_3' = p_3q_3
\end{align*}
where prime symbol means new values for the corresponding bits. Basically each line swaps two bits if \fm{p_i\!=\!0} and \fm{q_i\!=\!1} and leaves unchanged in all other cases. This operation does not change the value \fm{P\!+\!Q}.

{SCU} computes ancillary bits (such as \fm{p_4q_4}, \fm{p_3q_3})
and triggers a rule from the following logic:
\begin{empheq}[box=\fbox]{align*}
& \mathbf{if} && (p_4 q_4) && \mathbf{1:} && \mathbf{add}\ R_1 && - \\
& \mathbf{else\ if} && (p_4p_3q_3) && \mathbf{2:} && \mathbf{add}\ R_1 && p_4\!=\!q_4\!=\!0 \\
& \mathbf{else\ if} && (p_4) &&  \mathbf{3:} &&  \mathbf{add}\ R_n && p_4\!=\!0  \\ 
& \mathbf{else\ if} && (p_3q_3) && \mathbf{4:}  &&  \mathbf{add}\ R_n && q_4\!=\!0\\
& \mathbf{else} && && \mathbf{done} 
\end{empheq}
First, \fm{p_4q_4} is tested and if true {rule~1} is triggered. Its action is to add $R_1$ to the accumulator. The addition is done in \fm{n\!+\!1} size carry-save adder. In this case discarded overflow is automatically balanced with the addition of $R_1$. 
If rule~1 is not triggered, the condition of rule~2 is tested and if triggered, $R_1$ is added and then bits $p_4$ and $q_4$ are cleared (set to zero). If not, we proceed to the next clause. 
Rules~3 and 4 work similarly.
If any rule is triggered we cycle back to the beginning. 
Rule 1 or 2 can be triggered only in the first iteration because in the subsequent iterations $q_4$ and \fm{(p_4p_3q_3)} cannot be one. 

The idea behind this logic is simple: we keep subtracting power of 2: $\beta$ or $2\beta$, and compensate by adding $R_n$ or $R_1$ until the accumulator value is reduced.
Subtraction is done by either overflowing (rule 1) or directly clearing the bits after addition.

\input{diag5_sh}

%% file: diag5_sh.tex
\begin{figure}[t]

\begin{center}

\begin{tikzpicture}[auto, node distance=1.5cm,>=latex']
\node [point] (input){};
\node [topup, below of=input, node distance=1cm] (start) {top-up};
\node [none, below of=start] (scuC){};
\node [none, right of=scuC, node distance=1.25cm] (scuRm){};
\node [none, right of=scuRm, node distance=1.25cm] (scuR){};
\node [none, below of=scuR, node distance=0.06cm] (scuRd){};
\node [block, left of=scuC, node distance=1.25cm] (scu){SCU};

\node [rule, below of=scuC] (rule1){rule 1};
\node [rule, below of=rule1](rule2){rule 2};
\node [rule, below of=rule2](rule3){rule 3};
\node [rule, below of=rule3](rule4){rule 4};

\draw [darr,draw=none] (rule1) -- node[name=rul12] {} (rule2);

\node [sum, right of=rul12, node distance=1.5cm] (sum1){\fm{\Sigma{R_1}}};
\node [sum, below of=sum1, node distance=1.5cm] (sum2){\fm{\Sigma{R_n}}};

\node [none, below of=scu, node distance=3.25cm] (end) {};
\node [none, below of=end, name=out] {};

\draw [darr] (input) -- node {$|P,Q|\!=\!n\!+\!1$} (start);
\draw [darr] (start) -| (scu);
\draw [darr] (end) -- node {$|P,Q|\!=\!n$}(out);

\draw [darr] (scu) |- (rule1);
\draw [darr] (scu) |- (rule2);
\draw [darr] (scu) |- (rule3);
\draw [darr] (scu) |- (rule4);
\draw [darr] (scu) -- (out);%
\draw [darr] (scuRd) |- node[anchor=west, near start]{$0..3$}(start);
\draw [darr] (rule1) -- (sum1);
\draw [darr] (rule2) -- (sum1);
\draw [darr] (rule3) -- (sum2);
\draw [darr] (rule4) -- (sum2);
\draw [darr] (sum1) -| (scuR);
\draw [darr] (sum2) -| (scuR);

\end{tikzpicture}

\end{center}

\caption{Shrink module executes up to three cycles applying one rule at a time determined by SCU (Shrink Control Unit). Its task is to clear the top bits of $P$ and $Q$.}
\label{f:diag_sh}
\end{figure}
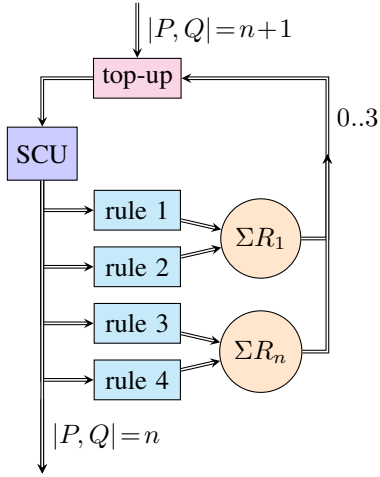

%% file: 45-cyc.tex
\subsection{Number of cycles}
Number of Shrink cycles cannot be less than three because of the counterexample
\mbox{$n\!=\!8$}, \mbox{$A\!=\!63$}, \mbox{$B\!=\!121$}, \mbox{$R\!=\!173$}.
On the other hand it is obvious that the number is not greater than 7, since \fm{P\!+\!Q\!<\!4\beta} and every iteration subtracts a value greater than \fm{\beta/2}.
It turns out proving that {\it the number of Shrink cycles is not greater than 4} is not too difficult. 

\noindent {\bf Proof:} Consider three possible cases:
\begin{enumerate}
    \item One of the top bit of $P$ or $Q$ is zero;
    \item Both top bits of $P$ and $Q$ are ones {\it and} \mbox{$R<R_c$}; and
    \item Both top bits of $P$ and $Q$ are ones {\it and} \mbox{$R>R_c$}.
\end{enumerate}
Here $R_c$ is a critical value for $R_1$ \mbox{$(=\![2\beta]_R)$} such that
\begin{align*}
& R_1 = 2\beta - 2R - \alpha R \\
& \alpha = 1  \quad \mathbf{if} \quad R<R_c \\
& \alpha = 0  \quad \mathbf{if} \quad R>R_c 
\end{align*}
Therefore
\begin{equation*}
R_c = \frac{2}{3}\beta
\end{equation*}
Remember that \mbox{$\beta=2^n$}.
Note, that $R$ is never equal to $R_c$. The above three cases are exhaustive, thus it is sufficient to prove each case.

\noindent {\bf Case 1:} Since one of the top bits of $P$ or $Q$ is zero, the total value is limited by \mbox{$P\!+\!Q<3\beta$}. Each cycle reduces the accumulator by $\beta$ and adds $R_n$. After four cycles the new accumulator values $(P',Q')$ are
\begin{align*}
P'+Q'=P+Q+4(-\beta+R_n)
\end{align*}
Since \fm{R_n\!=\![\beta]_R} and \fm{R\!\ge \!\beta/2}, $R_n$ can be either zero: \mbox{$R_n\!=\!0$} if \fm{R\!=\!\beta/2}, or if \fm{R\!>\!\beta/2}, then:
\begin{align*}
&    R_n=\beta-R \quad \mathbf{and} \quad R_n < \frac{1}{2}\beta\\
&P'+Q'< 3\beta +4(-\beta+\frac{1}{2}\beta) = \beta
\end{align*}
The final result is less than $\beta$ $(=2^n)$. Hence both $P'$ and $Q'$ each less than $\beta$, their top bits are zero \fm{p_4\!=\!q_4\!=\!0} and \fm{p_3q_3\!=\!0}.

\noindent{}{\bf Case 2:} If both top bits of $P$ and $Q$ are ones, then the first cycle reduces the accumulator by $2\beta$ and adds $R_1$; and the following three cycles do the same as in Case 1:
\begin{equation*}
\begin{split}
P'+Q'& = P+Q+(-2\beta+R_1)+3(-\beta+R_n)=\\
 & = P+Q - 5\beta + R_1 +3R_n
\end{split}    
\end{equation*}
Since now \mbox{$P\!+\!Q<4\beta$}, it would be sufficient to prove that \mbox{$R_1\!+\!3R_n<2\beta$}. In Case 2 \mbox{$R<R_c$} and \mbox{$\alpha=1$}, so
\begin{equation*}
\begin{split}
R_1 +3R_n &= (2\beta -3R)+3(\beta - R) = \\
& = 5\beta -6R \;\; < \;\; 5\beta - 6\cdot \frac{1}{2}\beta = 2\beta
\end{split}    
\end{equation*}
This proves that \mbox{$R_1\!+\!3R_n<2\beta$}, therefore 
\mbox{$P'\!+\!Q'<\beta$}, hence \fm{p_4\!=\!q_4\!=\!0} and \fm{p_3q_3\!=\!0}

\noindent{\bf Case 3:} When \fm{R>R_c}, the same logic follows as in Case 2 up to the derivation of  \fm{R_1\!+\!3R_n}. This time, however, \fm{\alpha=0} and $R$ is bound by \fm{R_c=2\beta/3} from below:
\begin{equation*}
\begin{split}
R_1 +3R_n &= (2\beta -2R)+3(\beta - R) = \\
& = 5\beta -5R \;\; < \;\; 5\beta - 5\cdot \frac{2}{3}\beta = \frac{5}{3}\beta < 2\beta
\end{split}    
\end{equation*}
As above in Case 2 this proves that \mbox{$R_1\!+\!3R_n<2\beta$}, therefore 
\mbox{$P'\!+\!Q'<\beta$}, hence \fm{p_4\!=\!q_4\!=\!0} and \fm{p_3q_3\!=\!0}.
$\square$

In the three cases above it was assumed \fm{P\!+\!Q\!<\!4\beta} at the beginning. In reality the upper bound is lower, since there are dependencies between the bits of $P$ and $Q$. For example, $P$ and $Q$ cannot both be equal to \fm{2\beta-1} upon the exit from the main loop. 
It is likely that the number of Shrink cycles is not greater than 3.
Proving this statement is much harder. It might be possible to prove formally using symbolic execution or {\it Binary Decision Diagrams}. At the time of writing, the idea of the proof may go along the following arguments. The values $R_x$ have at least one zero in the top three bits. This is a result of binary expansions of $2/3$, $4/5$, and $6/7$ - some of the critical values for $R_1, R_2, R_3$. This causes, after the main loop, having at least one zero in the three top bits of $P$ and $Q$. The reduction in each cycle is done faster because \fm{P\!+\!Q} value is smaller and \fm{\beta\!-\!R_n} (and \fm{2\beta\!-\!R_1}) is greater. The statement that 3 cycles are sufficient is left as a conjecture.

%% file: 50-squee.tex
\section{Squeeze module}
\input{diag6_sq}

The output of Shrink module are $P$ and $Q$ value of size $n$ (i.e. \fm{p_4\!=\!q_4\!=\!0}) and \fm{p_3q_3\!=\!0}.
Here again, as above, we use 4-based indices as in the example in Section \ref{ss:lcu} instead of $n$-based.
Squeeze module, depicted in \fig{f:diag_sq}, reduces further the accumulator to make both $P$ and $Q$ be less than $R$. Its work is similar to Shrink's except that it does not have cycles - one of its six rules is applied once. 
Squeezer starts with {Top-up}, which is similar to Shrink's but this time it operates on the two next to the top bits:
\begin{align*}
    &p_3' = p_3\spar{}q_3 & &q_3' = p_4q_3 \\
    &p_2' = p_2\spar{}q_2 & &q_2' = p_2q_2
\end{align*}
{\it S{\bf\emph{q}}ueezer Control Unit} ({QCU}) analyses top bits of the accumulator and makes decision on which rule to apply using the following logic:
\begin{empheq}[box=\fbox]{align*}
& \mathbf{if} && (\overline{p}_3) && \mathbf{1:}  &&  \qquad - && \\
& \mathbf{else\ if} && (q_2) && \mathbf{2:}  && p_3\!=\!p_2\!=\!q_2\!=\!0 && \mathbf{add}\ R_n\\
& \mathbf{else\ if} && (\overline{r}_2p_2) && \mathbf{3:}  && p_3\!=\!p_2\!=\!0 && \mathbf{add}\ R_m\\
& \mathbf{else\ if} && (\overline{r}_2\overline{p}_2) && \mathbf{4:}  &&  p_2\!=\!q_2\!=\!1,\ p_3\!=0\mkern-22mu &&  \\
& \mathbf{else\ if} && (r_2\overline{p}_2) && \mathbf{5:}  && \qquad - && \\
& \mathbf{else\ if} && ({r}_2p_2) && \mathbf{6:}  && p_2\!=0,\ q_2\!=\!1 &&
\end{empheq}
Note, that the order of clearing bits and summation is opposite comparing to Shrink's SCU. 
Along with accumulator bits QCU also inspects the second top bit of $R$, $r_2$. Depending on its value either rules (3,~4) or (5,~6) work.


{\bf Rule~1} says that if $p_3$ is zero, then we are done because $q_3$ is zero and $r_3$ is always one, ensuring the exit condition.
{\bf Rule~2} is triggered when $p_3,p_2,q_2$ bits are set. It subtracts $\beta$ by clearing these bits and compensates with $R_n$.
Now if \fm{R\!<\!3\beta/4} (i.e. \fm{r_2=0}) rules~3 and 4 are active.
{\bf Rule~3} makes $R_m$ reductions.
And {\bf rule~4} does not reduce the accumulator. To ensure the exit condition, it subtracts \fm{\beta/2} from $P$ and adds \fm{\beta/4} to $P$ and $Q$, so \fm{P\!+\!Q} remains the same. Now \fm{P\!<\!R} because \fm{r_3\!=\!1} but \fm{p_3\!=\!0}; same for $Q$. 
For {\bf rule~5} we are done, because \fm{p_2\!=\!0} implies \fm{q_2\!=\!0} but \fm{r_2\!=\!1}. {\bf Rule~6} does the trick similar to rule 4: subtracting \fm{\beta/4} from $P$ and adding to $Q$. This works because here \fm{R\!\ge \!3\beta/4} and, before the rule is executed, \fm{P\!<\!\beta} and \fm{Q\!<\!\beta/4}. 
The conditions of these rules list all possible combinations of the accumulator values. The detailed inspection of each rule proves that the result satisfies the exit conditions.
Note, that rules missing addition should not perform addition with zero because carry-save addition of $P$ and $Q$ (even with zero) changes their values and may change the top bits.

%% file: diag6_sq.tex
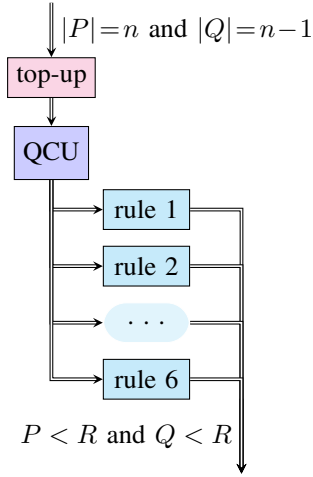
\begin{figure}[t]

\begin{center}

\begin{tikzpicture}[auto, node distance=1.5cm,>=latex']
\node [point] (inputC){};
\node [none, below of=inputC, node distance=1cm] (startC) {};
\node [none, left of=inputC, node distance=1.25cm] (input) {};
\node [topup, left of=startC, node distance=1.25cm] (start) {top-up};
\node [none, below of=startC] (scuC){};
\node [none, right of=scuC, node distance=1.25cm] (scuR){};
\node [none, below of=scuR, node distance=0.06cm] (scuRd){};
\node [block, left of=scuC, node distance=1.25cm] (scu){QCU};

\node [rule, below of=scuC] (rule1){rule 1};
\node [rule, below of=rule1](rule2){rule 2};
\node [ruleDots, below of=rule2](rule3){ . . . };
\node [rule, below of=rule3](rule4){rule 6};

\node [none, below of=scuR, node distance=3.25cm] (end) {};
\node [point, below of=end, name=out] {};

\draw [darr] (input) -- node {$|P|\!=\!n$ and $|Q|\!=\!n\!-\!1$} (start);
\draw [darr] (start) -- (scu);
\draw [darr] (end) -- node [anchor=east]{$P<R$ and $Q<R$}(out);

\draw [darr] (scu) |- (rule1);
\draw [darr] (scu) |- (rule2);
\draw [darr] (scu) |- (rule3);
\draw [darr] (scu) |- (rule4);
\draw [darr] (rule1) -| (out);
\draw [darr] (rule2) -| (out);
\draw [darr] (rule3) -| (out);
\draw [darr] (rule4) -| (out);

\end{tikzpicture}

\end{center}

\caption{Squeeze module applies one rule out of six reducing the accumulator to its final state.}
\label{f:diag_sq}
\end{figure}